\documentclass[twocolumn,showpacs,preprintnumbers,amsmath,amssymb,aps]{revtex4}

\usepackage{amssymb}
\usepackage{amsmath,amssymb,graphicx}
\usepackage{graphicx}
\usepackage{dcolumn}
\usepackage{bm}
\def\be{\begin{equation}}
\def\ee{\end{equation}}
\def\bea{\begin{eqnarray}}
\def\eea{\end{eqnarray}}

\def\lsim{\raise0.3ex\hbox{$\;<$\kern-0.75em\raise-1.1ex\hbox{$\sim\;$}}}
\def\gsim{\raise0.3ex\hbox{$\;>$\kern-0.75em\raise-1.1ex\hbox{$\sim\;$}}}

\begin{document}

\title{A new strategy for probing  the  Majorana neutrino CP violating phases and masses}%

\author{David Delepine}
\affiliation{Departamento de Fisica,  Universidad de
Guanajuato, Campus Leon,
 C.P. 37150, Le\'on, Guanajuato, M\'exico.}%
\author{Vannia Gonz\'alez Mac\'\i as}
\affiliation{Departamento de Fisica,  Universidad de
Guanajuato, Campus Leon,
 C.P. 37150, Le\'on, Guanajuato, M\'exico.}%
\author{Shaaban Khalil}
\affiliation{Centre for Theoretical Physics, The British
University in Egypt, El Sherouk City,
Postal No, 11837, P.O. Box 43, Egypt}%
\author{Gabriel L\'opez Castro}
\affiliation{Departamento de Fisica, Cinvestav, Apartado Postal 14-740, 07000 Mexico D.F., Mexico}%

\date{\today}

\begin{abstract}
We propose a new strategy for detecting  the  CP-violating
phases and the effective mass of muon Majorana neutrinos by measuring  observables associated with neutrino-antineutrino oscillations in $\pi^{\pm}$ decays. Within the generic framework of quantum field theory, we compute the non-factorizable probability for producing a pair of same-charged muons in $\pi^{\pm}$ decays as a distinctive signature of $\nu_{\mu}-\bar{\nu_{\mu}}$ oscillations. We show that an intense neutrino beam through a long
baseline experiment is favored for probing the Majorana phases.
Using the neutrino-antineutrino oscillation probability reported
by MINOS collaboration, a new stringent bound on the effective
muon-neutrino mass is derived.

\end{abstract}

\pacs{11.30.Er,11.30.Hv, 13.15.+g, 14.60.Pq}
\maketitle

\section{Introduction.}
Detecting CP violating phases in the lepton sector is one of the
most challenging problems in the study of neutrino mixing. In the
basis of diagonal charged lepton mass matrix, the neutrino mass
matrix $m_{\nu}$ can be written in flavor basis as $m_{\nu} = U^*
m_{\nu}^{{\rm diag}} U^{\dag}$, where $U$ is the PMNS mixing
matrix, which can be written as $U=V.{\rm diag}\left(1,
e^{i\alpha_2}, e^{i\alpha_3}\right)$. $\alpha_i$ are the Majorana
phases and the mixing matrix $V$ can be parameterized by one Dirac
phase and three angles: solar angle $\theta_{12}$, atmospheric
angle $\theta_{23}$, and Chooz angle $\theta_{13}$.

The measurement of the effective electron-neutrino mass in the neutrinoless
double beta decay ($0\nu\beta\beta$) experiments can not restrict
the two Majorana CP violating phases present in the PMNS mixing
matrix. The effective electron-neutrino mass $\langle m_{ee} \rangle$ is given
by
\begin{eqnarray}%
\left\vert \langle m_{ee}\rangle \right\vert &=& \Big\vert \sum_i
U_{ei}^2  m_{\nu_i} \Big\vert \nonumber\\ %
&=& \Big\vert m_{\nu_1}  U_{e1}^2 + m_{\nu_2} U_{e2}^2
+ m_{\nu_3} U_{e3}^2  \Big\vert
~.~~~
\end{eqnarray}%

This effective mass parameter depends on  the angles
$\theta_{12}$ and $\theta_{13}$, the neutrino masses $m_{\nu_i}$, Dirac
CP phase, and Majorana phases $\alpha_i$. There are several
studies on using the results of ($0\nu\beta\beta$) together with
the new data from terrestrial and astrophysical observation in
order to restrict the Majorana neutrino CP violating phases\cite{betadecay,2,3}.
However, this analysis is model dependent and quite sensitive to
the ansatz of the neutrino mass spectrum: quasi-degenerate, normal
or inverted hierarchies. In this respect, it is not possible to
measure the Majorana neutrino CP phases from ($0\nu\beta\beta$)
experiment. This may be expected since in the ($0\nu\beta\beta$)
one measures the lifetime of the decay of two neutrons in a
nucleus into two protons and two electrons, which is a CP
conserving quantity.

On the other hand, direct bounds on other effective neutrino mass
parameters $\langle m_{ll}\rangle \equiv \sum_i U_{li}^2 m_{\nu_i}$ from present experimental data
are very poor. Currently, the strongest bound for the
muon-neutrino case from the $K^+ \to \pi^-\mu^+\mu^+$ branching
fraction \cite{kmumu} is only $\langle m_{\mu\mu}\rangle \leq 0.04$ TeV
\cite{mll}, which leads \textbf{to a} negligible constraint on the
neutrino masses and CP violating phases. Therefore, it is commonly
believed that direct bounds from other $\Delta L=2$ decays are
only of academic interest and can not fix the neutrino mixing
parameters \cite{Merle:2006du}. Some attempts to detect CP
violation based on the difference between oscillation
probabilities of neutrinos and antineutrinos can be found in
Ref.\cite{mnp}.

Here we propose a mechanism, based on neutrino-antineutrino
oscillation
\cite{Bahcall:1978jn,Schechter:1980gk,Langacker:1998pv}, which
would allow to derive a strong bound on the effective mass of the
muon-neutrino. In addition, it provides a method for detecting the
Majorana neutrino CP violating phases through measuring the CP
asymmetry of the $\pi^{\pm}$ decay where neutrino-antineutrino
oscillation take place. Using the bound on the
neutrino-antineutrino oscillation probability reported by the
MINOS Collaboration \cite{hartnell}, we derive a bound on $\langle
m_{\mu\mu}\rangle$ which improves existing bounds by several
orders of magnitude.

It is worth noting that the probability of a process associated to
neutrino oscillation is usually assumed to be factorized into
three independent parts: the production process, the oscillation
probability and the detection cross section. In Ref. \cite{nos},
this approximation was avoided and a generic framework based on
quantum field theory was proposed to get a simple expression for
the CP asymmetry. Here, we adopt the S-matrix amplitude method
described in \cite{nos}.

\section{Neutrino-Antineutrino Oscillation}

 Let us start by considering a positive charged pion which
decays into a virtual neutrino at the space-time location
$(x,t)$ together with a  positive charged muon.  After propagating,
the neutrino can be converted into an antineutrino which produces a
positive charged muon at the point $(x^{'},t^{'})$ when it interacts with a target, as shown in Fig. 1.  For definiteness, we illustrate this process with
the production of the neutrino in $\pi^+$ decay and its later
detection via its weak interaction with a target nucleon $N$
\begin{eqnarray}
\pi^+ (p_1)\rightarrow & \mu^{+}(p_2)&+ \nu_{\mu}^s(p) \nonumber \\
&\hookrightarrow & \overline{\nu_{l}}^d(p) + N(p_N) \rightarrow N'(p_{N^{'}})+\mu^{+}(p_l)\ . \nonumber
\end{eqnarray}
 where the superscript $s(d)$ refers to the virtual neutrino (antineutrino) at the source (detection) vertex. This  $|\Delta L|=2$ process is a clear signal for neutrino-antineutrino oscillations of the muon type and its amplitude should be proportionnal to neutrino Majorana masses.

 If one ignores other flavors, the time evolution of the $\nu_{\mu}-\bar{\nu}_{\mu}$ system would be analogous to that of the $K^0-\bar{K}^0$ or $B^0-\bar{B}^0$ systems.
\begin{figure}
  \includegraphics[width=6cm]{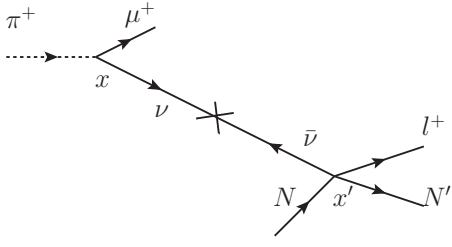}\\
  \caption{Feynman diagram of the process $\pi (p_1)\rightarrow
\mu^{+}(p_2)+ \nu_{\mu}(p)$ followed by the detection
process:$\overline{\nu_{l}}(p) + N(p_N) \rightarrow
N'(p_{N^{'}})+\mu^{+}(p_l)$.}\label{fig}
\end{figure}
 Instead, we prefer to use the formalism developed in Ref.\cite{nos}, where the whole reaction includes the production and detection processes of neutrinos. The decay amplitude becomes (for simplicity we assume that leptonic flavor is conserved at the production and detection vertices):
\begin{eqnarray}
T_{\nu_{\mu}-\bar{\nu}_{\mu}}(t)&=& (2 \pi)^4 \delta^4(p_l-p_N+p_{N^{'}}+p_2-p_1) \nonumber \\
&&\times (G_F V_{ud})^2 (J_{NN^{'}})_{\mu} f_{\pi} \nonumber \\
&&\times \sum_i \bar{v}_{\mu}(p_l) \gamma^{\mu}(1+\gamma_5) \not{p}_{1}v(p_2) \nonumber \\
&& \times U_{\mu i}U_{\mu i}(m_{\nu_{i}})\frac{e^{-itE_{\nu_i}}}{2E_{\nu_i}}\ ,
\end{eqnarray}
where the relation $\nu_k=\sum U_{k\alpha} \nu_{\alpha}$ between
flavor $k$ and mass $\alpha$ neutrino eigenstates has been used, $f_{\pi}=130.4$ MeV is the $\pi^{\pm}$ decay constant,
and $J_{NN^{'}}$ parametrizes the interaction with the nucleon.
 Note that, contrary to the case of neutrino oscillations \cite{nos}, only the neutrino mass term survives in this case. For simplicity, one assumes that
\begin{equation}
(J_{NN^{'}})_{\mu}=\overline{u}_{N^{'}}(p_{N^{'}})\gamma_{\mu}(g_V+g_A \gamma_5) u_{N}(p_{N})
\end{equation}
where we use $g_V=g_V(q^2=0)=1$ and $g_A=g_A(q^2=0)\approx-1.27$~
\cite{Yao:2006px}.  If we neglect  terms of $O(m_{\mu}/m_{N,N^{'}})$, one
obtains
\begin{eqnarray}
\left|T_{\nu_{\mu}-\bar{\nu}_{\mu}}(t)\right|^{2}&=&(2 \pi)^4 \delta^4(p_l-p_N+p_{N^{'}}+p_2-p_1)
(G_F V_{ud})^4 \nonumber \\
&\times& \left|f_{\pi}\right|^{2} \sum_{i,j} U_{\mu i}U_{\mu j}^{*}U_{\mu i}U_{\mu j}^{*}
e^{-it\triangle E_{\nu_{ij}}}\frac{m_{\nu_{i}}m_{\nu_{j}}}{4E_{\nu_{i}}E_{\nu_{j}}} \nonumber \\
&\times& 64 (g_{A}-1)^{2} m_{N} m^{2}_{\mu} (E_{2}-E_{p})
\nonumber \\
&\times& \left( \left[ 1 - \frac{m_{N}}{(E_{2}-E_{p})}
G(g_{A})\right]p_{l}\cdot
p_{2}\right.\nonumber \\
&-& \!\left. \!2m_{N}F(g_{A})\left[E_{2}
-E_{l}\left(1+\frac{m_{\pi}^{2}}{m_{\mu}^{2}}\right)\right]\right. \nonumber \\
&-&\! \left. \!\frac{1}{2}(m_{\mu}^{2}-\!m^{2}_{\pi})\right)
\label{tsquare}
\end{eqnarray}
where  $E_{2}(E_{l}),E_{p}$ are, respectively, the initial
(final) muon and the pion energies and $\triangle E_{\nu_{ij}}=E_{\nu_i}-E_{\nu_j}$. The functions $F(g_{A})$ and
$G(g_{A})$ are given by:
$F(g_{A})=\frac{g_{A}^{2}+1}{(g_{A}-1)^{2}}$,
$G(g_{A})=\frac{g_{A}+1}{g_{A}-1}$.  One can easily check that Eq. (\ref{tsquare}) is not factorizable into (production)$\times$(propagation)$\times$(detection) subprocesses due to the terms proportional to $p_{l}\cdot
p_{2}=E_{l}E_{2}-|\vec{p_{l}}||\vec{p_{2}}|\cos\alpha$,  where $\alpha$ is the angle between the directions of $\mu^+$ particles. This is an important difference with respect to the case of neutrino-neutrino ($\Delta L=0$) oscillations where it was
shown in Ref.\cite{nos} that the S-matrix formalism reproduces the hypothesis of factorization of the probabilities.

After integration over kinematical variables, it is possible to write the
rate of the complete process as
%
%
\begin{equation}
\Gamma_{\nu_{\mu}-\bar{\nu}_{\mu}}=  \left|\sum_{i} U_{\mu i}^2
\frac{m_{\nu_{i}}}{2E_{\nu_{i}}}e^{itE_{\nu_{i}}} \right|^2 \times
F(M,\phi),
\end{equation}
where $F(M, \phi)$ denotes the kinematical function
\begin{eqnarray}
F(M,\phi)&=& \frac{\pi}{2E_{p}}(G_F
V_{ud})^4\left|f_{\pi}\right|^{2}64(g_{A}-1)^{2}
\nonumber \\
&\times& \!\!\!
\left(\left[I_{4}-m_{N}G(g_{A})I_{5}-\frac{1}{2}(m_{\mu}^2-m_{\pi}^2)I_{1})
\right]m^{2}_{\mu}\frac{}{}\right.  \nonumber \\
&-& \left.
\!\!\!2m_NF(g_A)\left[m_{\mu}^2I_2-(m^{2}_{\mu}+m^{2}_{\pi})I_{3}\right]\frac{}{}\!\right).
\end{eqnarray}
The functions $I_{a}$ for $a=1,..,4$ can be obtained from the
following integral: %
\begin{equation}
I_{a}=\!\!\int \! \frac{d^{3}p_{2}}{2E_{2}}\frac{d^{3}p_{l}}{2E_{l}}
(E_{2}-E_{p})f_a \delta (E_{p}+E_{N}-E_{N'}-E_{l}-E_{2}),
\end{equation}
with $f_1=1$, $f_2 =E_{2}$, $f_3=E_{l}$, and $f_4=(p_{l}\cdot
p_{2})$ and \textbf{$f_{5}=\frac{(p_{l}\cdot
p_{2})}{E_{2}-E_{p}}$}.

There are two interesting limits to this process. At very
short times, which means that the detection is very close to the
production vertex (short-baseline neutrino experiment), one has,
 assuming that the $E_{\nu_{i}}\simeq E_{\nu}$, that
\begin{equation}
 \Gamma_{\nu_{\mu}-\bar{\nu}_{\mu}} \simeq \frac{|\langle m_{\mu \mu}\rangle |^2}{E_{\nu}^2} \times F(M,\phi)
 \end{equation}
where $\langle m_{\mu \mu}\rangle$ is the effective Majorana mass for the muon
neutrino. In the long time limit  which corresponds to a
long-baseline neutrino experiment, the oscillation terms cannot
be neglected and this process depends on a  new combination of
phases, mixing angles and masses which could give us complementary
information on the neutrinoless double beta decays or on any process
that depends exclusively on the effective Majorana mass of the
neutrinos. Using this expression for the rate it is possible to
get the CP asymmetry  which will depend explicitly on Majorana
phases.
\begin{eqnarray}
a_{CP}&=&\frac{\Gamma_{\nu_{\mu}-\bar{\nu}_{\mu}}-\overline{\Gamma}_{\nu_{\mu}-
\bar{\nu}_{\mu}}}{\Gamma_{\nu_{\mu}-\bar{\nu}_{\mu}}+\overline{\Gamma}_{\nu_{\mu}-\bar{\nu}_{\mu}}} \\
&=&\frac{\sum_{i>j}{\rm Im}\left(U_{ \mu i}U_{\mu i}U_{\mu
j}^{*}U_{\mu j}^{*}\right)m_{\nu_i}m_{\nu_j}\sin \gamma} {\sum_{i>j}{\rm
Re}\left( U_{\mu i}U_{\mu i}U_{\mu j}^{*}U_{\mu
j}^{*}\right)m_{\nu_i}m_{\nu_j}\cos \gamma} \label{acp}
\end{eqnarray}
where $\gamma=\frac{\Delta m_{23}^2 L(km)}{2 E_{\nu}(GeV)}$. Here
$\Delta m_{23}^2$ is the difference in the squares of second and third eigenstate neutrino masses, $\Delta
m_{23}^2=(2.43\pm 0.13) \times 10^{-3}$ eV$^2$, and $L$ is the
distance between production and detection vertices. Finally, $E_{\nu}$ is the
energy of the neutrino beam.  It is worth mentioning that the
time evolution amplitude for the CP-conjugate process corresponds
to  the observation of $\mu^-$ at the source  and at the
detector. Therefore, the associated nucleon weak vertex is given
by $(J_{N^{'}N})_{\mu}=\overline{u}_{N}(p_{N})\gamma_{\mu}(g_V+g_A
\gamma_5) u_{N^{'}}(p_{N^{'}})$. In~estimating the CP asymmetry in
Eq.(\ref{acp}), we have assumed that $J_{N^{'}N} \simeq
J_{NN^{'}}$.

In the limit of $\theta_{13}=0$, the Majorana phases
$\alpha_{1,2}$ are the only sources of CP violation and
hence ${\rm Im}(U_{\mu i} U_{\mu i} U^*_{\mu j} U^*_{\mu j})
\propto \sin
(\alpha_i -\alpha_j)$. For $i=2$ and $j=3$ one finds%
\begin{equation}%
a_{CP} \simeq \tan\left[2(\alpha_2 - \alpha_3)\right] \sin \gamma.
\end{equation}
Thus, in the case of long-baseline neutrino experiment like MINOS
where the distance $L$ is given by $L=735$ km and the energy
$E_{\nu}$ is typically around $2-3$ GeV
\cite{Arms:2009zz,Blake:2008zz}, one finds that $\sin \gamma \sim
{\cal O}(1)$. Thus, measuring CP asymmetry  will
be unavoidable indication for large CP violating Majorana phases.

\section{ Application to MINOS results on neutrino-antineutrino oscillations}
 The last two decades have witnessed several experiments that
investigate the neutrino-antineutrino transitions. It started
in 1982 when the BEBC bubble chamber in the CERN SPS neutrino beam
set a limit on $\nu_{\mu} \to \bar{\nu}_e$ and $\nu_{e} \to
\bar{\nu}_e$ through the search for $\bar{\nu}_e$ appearance.
Recently,  MINOS \cite{Blake:2008zz} has measured the spectrum of $\nu_{\mu}$
events which are missing after travelling $735$ km. It is these missing
events which are the potential source of $\bar{\nu}_{\mu}$
appearance. In their preliminary analysis, they were able to put a
limit on the fraction of muon neutrinos  transition to  muon
anti-neutrinos \cite{hartnell}:
\begin{equation}
P(\nu_{\mu}-\bar{\nu}_{\mu}) < 0.026~  (90 \% \ {\rm c.l.}).
\end{equation}

Assuming CPT, this limit can be written as
\begin{eqnarray}
\frac{\Gamma_{\nu_{\mu}-\bar{\nu}_{\mu}}}{\Gamma_{\nu_{\mu}-\nu_{\mu}}}&<& 0.026\ .
\end{eqnarray}
Using our expression for $\Gamma_{\nu_{\mu}-\bar{\nu}_{\mu}}$, and the corresponding rate for neutrino oscillations \cite{nos}, one gets
\begin{eqnarray}
\left|\sum_{i} U_{\mu i}^2
\frac{m_{\nu_i}}{E_{\nu_i}}e^{itE_{\nu_{i}}} \right|^2 &
\lsim& 0.001 \ .
\end{eqnarray}
In the limit of ultrarelativistic neutrinos, $E_{\nu_i} \simeq E_{\nu}(1+m_{\nu_i}^2/2E_{\nu})$. and keeping the leading terms in the $m_{\nu_i}/E_{\nu}$ terms, we get
\begin{equation}
\left|\sum_{i} U_{\mu i}^2 m_{\nu_i}e^{it\frac{ m^2_{\nu_i}}{ 2E_{\nu}}}\right|^2
\lsim 0.001 \times E_{\nu}^2 \ .
\end{equation}
To illustrate the usefulness of this relation, let us assume the case of two flavor neutrinos. In this case, one finds
\begin{eqnarray}
0.001 \times E_{\nu}^2 &\gsim &  \left|\langle m_{\mu \mu}\rangle \right|^2\nonumber \\
&-& 4~ {\rm Re} \left( U_{\mu 2}^2 U_{\mu 3}^{*2}\right)  m_{\nu_2} m_{\nu_3} \sin^2 \frac{\gamma}{2} \nonumber \\
&-& 2~ {\rm Im} \left( U_{\mu 2}^2 U_{\mu 3}^{*2}\right)m_{\nu_2} m_{\nu_3} \sin \gamma   \label{majorana}\ .
\end{eqnarray}

 From this equation, it is possible to get a bound on the effective muon-neutrino Majorana mass, only depending on the values of the Majorana phases as the oscillation terms cannot be neglected. If $\pi-\gamma/2 \leq 2(\alpha_2-\alpha_3) \leq 2\pi-\gamma /2$ and using $E_{\nu}\approx 2$ GeV,  one gets the following conservative bound on
$$
|\langle m_{\mu \mu}\rangle | \lsim 64 \ {\rm MeV}
$$
If   $-\gamma/2 \leq 2(\alpha_2-\alpha_3) \leq 0$, the conservative bound on  $|\langle m_{\mu \mu}\rangle |$ is given by
$$
|\langle m_{\mu \mu}\rangle | \lsim 109 \ {\rm MeV}
$$
If $ 0 \leq 2(\alpha_2-\alpha_3) \leq \pi-\gamma /2$,  it is not possible to get a conservative bound on $|\langle m_{\mu \mu}\rangle |$ but the equation (\ref{majorana}) could be used to bound Majorana parameters (masses and phases) which appear in $|\langle m_{\mu \mu}\rangle |$.


 So, within the hypothesis done on Majorana phases,
the limits obtained improve by various order of magnitude the actual limit
on $\langle m_{\mu \mu}\rangle $ coming from direct search in  $K^+
\to \pi^-\mu^+\mu^+$ decay. Also,  one could expect improvement in
experimental analysis in the next years and we could expect that
with a specially designed neutrino experiment to measure these kind
of processes, it should be possible to improve these limits by a
few orders of magnitude in future experiments.


In MINOS and in general in all long-baseline neutrino experiments,
the oscillation terms are not negligible. So, it means that such
analysis could not only give us information on the neutrino
effective Majorana mass  but it could be used to determine
parameters of the mixing matrices and bound on the absolute value
of Majorana masses. Also, if in a long-baseline neutrino
experiment, neutrino detectors are located at different distances
from the source, it should be possible to get enough constraints
on the mixing parameters and Majorana masses to fix them.

\section{Conclusions}

 If Majorana neutrinos do exist, $|\Delta L|=2$ processes like neutrino-antineutrino oscillations can occur. The production of  leptons with same charges at the production and detection vertices of neutrinos will be a clear manifestation of these processes. In this paper we have used the S-matrix formalism of quantum field theory to describe these oscillations {in the case of} muon neutrinos produced in $\pi^+$ decays  which convert into muon antineutrinos that are detected via inverse beta decay on nucleons.

One interesting result is that the time evolution probability of the whole process is not factorizable into production, oscillation and detection probabilities, as is the case in neutrino oscillations \cite{nos}. We find that, for very short times of propagation of neutrinos, the observation of $\mu^+\mu^+$ events would lead to a direct bound on the effective mass of muon Majorana neutrinos. In the case of long-baseline neutrino experiments, the CP rate asymmetry for production of $\mu^+\mu^+/\mu^-\mu^-$ events would  lead to direct bounds on the difference of CP-violating Majorana phases. Finally, using the current bound on muon neutrino-antineutrino oscillations reported by the MINOS Collaboration we are able to set the bound $\langle m_{\mu\mu} \rangle \lsim 64$ MeV, which is several orders of magnitude below current bounds reported in the literature.

  As a consequence of these results, neutrino experiments aiming to measure neutrino-antineutrino oscillations with different short- and long-baseline setups can be useful to get direct and complementary constraints on the masses and phases of Majorana neutrinos.

\section{acknowledgement}

This work was also partially supported by Conacyt (Mexico) and by
PROMEP project. The work of S.K. is supported in part by ICTP
project 30 and the Egyptian Academy of Scientific Research and
Technology. The authors would like to thank M. Bishai for her
discussions and attracting our attention to MINOS analysis.

\end{document}